\documentclass[aps, twocolumn, superscriptaddress,longbibliography, amsfonts,amssymb,amsmath,a4paper,floats,floatfix]{revtex4-2}

\usepackage[utf8]{inputenc}
\usepackage[english]{babel}

\usepackage{graphicx}
\usepackage{dcolumn}
\usepackage{bm}
\usepackage{siunitx}
\usepackage{xspace}
\usepackage{color}
\usepackage{braket}

\newcommand{\mos}{MoS\ensuremath{_{\mathrm{2}}}\xspace}
\newcommand{\vtg}{\ensuremath V_{\mathrm{TG}}}
\newcommand{\vbg}{\ensuremath V_{\mathrm{BG}}}   
\newcommand{\vsd}{\ensuremath V_{\mathrm{SD}}} 
\begin{document}

\title{Evidence of the Coulomb gap in the density of states of \mos}

\author{Michele Masseroni}%
    \affiliation{Solid State Physics Laboratory, ETH Z\"urich, 8093 Z\"urich, Switzerland}
\author{Tingyu Qu}%
    \affiliation{NUS Graduate School, Integrative Sciences and Engineering Programme (ISEP), National University of Singapore, Singapore 119077, Singapore}
\author{Takashi Taniguchi}
	\affiliation{International Center for Materials Nanoarchitectonics,  1-1 Namiki, Tsukuba 305-0044, Japan}
\author{Kenji Watanabe}
	\affiliation{Research Center for Functional Materials, 1-1 Namiki, Tsukuba 305-0044, Japan}
\author{Thomas\ Ihn}
    \affiliation{Solid State Physics Laboratory, ETH Z\"urich, 8093 Z\"urich, Switzerland}
\author{Klaus\ Ensslin}
    \affiliation{Solid State Physics Laboratory, ETH Z\"urich, 8093 Z\"urich, Switzerland}

\date{\today}

\begin{abstract}

$\mathrm{MoS_2}$ is an emergent van der Waals material that shows promising prospects in semiconductor industry and optoelectronic applications. However, its electronic properties are not yet fully understood. In particular, the nature of the insulating state at low carrier density deserves further investigation, as it is important for fundamental research and applications. In this study, we investigate the insulating state of a dual-gated exfoliated bilayer $\mathrm{MoS_2}$ field-effect transistor by performing magnetotransport experiments. We observe positive and non-saturating magnetoresistance, in a regime where only one band contributes to electron transport. 
At low electron density ($\sim\SI{1.4E12}{cm^{-2}}$) and a perpendicular magnetic field of 7 Tesla, the resistance exceeds by more than one order of magnitude the zero field resistance and exponentially drops with increasing temperature. We attribute this observation to strong electron localization. Both temperature and magnetic field dependence can, at least qualitatively, be described by the Efros-Shklovskii law, predicting the formation of a Coulomb gap in the density of states due to Coulomb interactions. However, the localization length obtained from fitting the temperature dependence exceeds by more than one order of magnitude the one obtained from the magnetic field dependence. We attribute this discrepancy to the presence of a nearby metallic gate, which provides electrostatic screening and thus reduces long-range Coulomb interactions.
The result of our study suggests that the insulating state of $\mathrm{MoS_2}$ originates from a combination of disorder-driven electron localization and Coulomb interactions.

\end{abstract}

\maketitle

\section{Introduction}

The resistivity $\rho$ of some semiconductors shows a metal-insulator transition as a function of the electron density $n$ \cite{imada_metal-insulator_1998}.
For densities larger than a critical value $n_\mathrm{c}$ the resistivity shows a metallic temperature dependence ($\mathrm{d}\rho/\mathrm{d} T>0$), while below $n_\mathrm{c}$ it shows an insulating temperature dependence ($\mathrm{d}\rho/\mathrm{d}T<0$). 
This metal-insulator transition attracted great interest in the late 1990s \cite{castellani_metallic_1998, phillips_superconductivity_1998, chakravarty_wigner_1999, klapwijk_few_1999}.
In two-dimensional (2D) semiconductors the origin of the metallic phase is controversial \cite{das_sarma_charged_1999, das_sarma_screening_1995, shashkin_metalinsulator_2021}, as it was predicted that any amount of defects would inexorably lead to electron localization at zero temperature in a 2D system \cite{anderson_absence_1958, abrahams_scaling_1979}.
The insulating phase at low densities can be due to either intriguing correlated states, like Wigner crystals \cite{shayegan_wigner_2022} or disorder-induced electron localization \cite{klapwijk_few_1999}, as well as a combination of the two effects.

In highly disordered systems, charge transport at low temperatures occurs via electron hopping between localized states \cite{shklovskii_electronic_1984}, known as variable-range hopping (VRH).
The conductivity in hopping transport at zero magnetic field is usually described by an exponential dependence on the temperature of the form
\[ \sigma(T) \propto \exp\left[-\left(\frac{T_0}{T}\right)^{p}\right], \]
where $T_0$ and $p\leq1$ are constants that depend on the hopping mechanism.
In a non-interacting system, the density of states close to the Fermi energy is constant (but finite) and the conductivity is described by Mott's law \cite{mott_conduction_1968}, for which $p=1/3$ (for two-dimensional systems).
When electrons are strongly localized, the long-range Coulomb potential is not efficiently screened.
Electron correlations result in a Coulomb gap in the density of states close to the Fermi energy \cite{pollak_effect_1970, ambegaokar_hopping_1971}. The modified density of states changes the temperature dependence of the hopping conductivity, which is now characterized by the parameter $p=1/2$, as described by the Efros-Shklovskii (ES) theory \cite{efros_coulomb_1975}. 

The insulating phase of \mos has been experimentally studied in monolayers \cite{radisavljevic_mobility_2013} and multilayers \cite{xue_mott_2019, moon_soft_2018}, where both thermally activated transport at intermediate temperatures ($T\sim \SIrange{50}{100}{K}$) and Mott VRH transport at lower temperatures have been observed. 
In addition, it is expected that electron-electron interactions play a major role in determining the electronic properties due
to the large electron effective mass [$m^*\approx (0.4-0.6)m_0$] of \mos, especially at low densities.
Indeed, signatures of interaction effects have already been reported in the literature \cite{lin_determining_2019, pisoni_interactions_2018}, among which there was also the observation of a Wigner crystal in MoSe$_2$ \cite{smolenski_signatures_2021}. 
Therefore, \mos, and in general, semiconducting transition metal dichalcogenides (TMDs), are good candidates for the observation of the Coulomb gap in the density of states. 
However, the observation of interaction effects is restricted to low densities, where the Coulomb energy dominates over the kinetic energy of electrons. 
Transport experiments in this density range are challenging in most materials and require low defect densities \cite{yoon_wigner_1999}.
The observation of the Coulomb gap in \mos remains to date elusive \cite{moon_soft_2018} due to the large density of intrinsic defects.

\begin{figure*}[tb]
    \centering
    \includegraphics{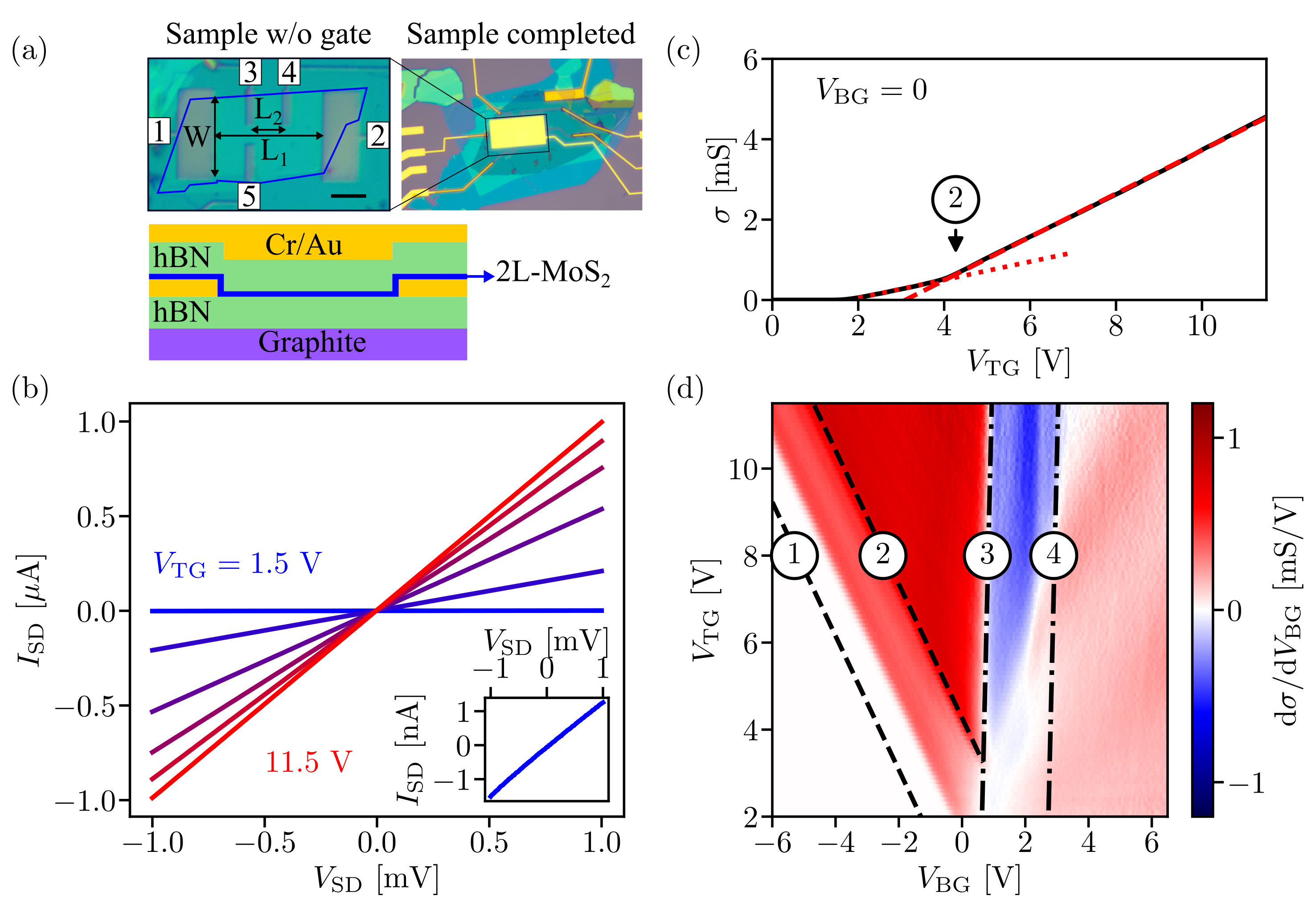}
    \caption{(a) The top panel shows optical micrographs of the sample before (left) and after (right) depositing the metallic top gate. The \mos flake is outlined in the left figure. The scale bar in the left figure is $\SI{4}{\mu m}$. The dimensions of the device are $W=\SI{8}{\mu m}$, $L_1=\SI{12}{\mu m}$, and $L_2=\SI{4}{\mu m}$. A schematic side view of the device is shown in the lower panel. (b) Current ($I_\mathrm{SD}$) versus voltage ($V_\mathrm{SD}$) characteristics for different $\vtg$ (step 2V). Inset shows the output characteristics for the lowest voltage $\vtg=\SI{1.5}{V}$.
    (c) Four-terminal conductivity ($\sigma$) as a function of top gate voltage ($\vtg$) at $\vbg=0$. The dashed and dotted lines are linear fits, which highlight a kink in $\sigma(\vtg)$ (marked by the number 2). 
    (d) Derivative of the conductivity $\mathrm{d}\sigma/\mathrm{d}\vbg$ as a function of top and bottom gate voltages. The black dashed lines follow the constant total density condition. 
    The dashed dotted lines mark a local maximum and minimum of the conductivity. The numbers 1,2,3,4 represent the band edges of the four bands (see discussion in Sec.~\ref{Sec:Band_and_layer_occupation}) that contribute to electron transport. }
    \label{fig:Fig1}
\end{figure*}

Here, we investigate magnetotransport in bilayer \mos encapsulated in hexagonal boron nitride (hBN). 
We first demonstrate the high quality of our device and
a complete understanding of the bands that contribute to electron transport. 
Then we tune the density, such that electron transport occurs in a single (twofold degenerate) band. 
In this regime, we observe a metal-insulator transition at the electron density $n_\mathrm{c}\approx \SI{1.7E12}{cm^{-2}}$. 
Below this transition, the zero-field resistance and the magnetoresistance show an exponential decay with increasing temperature, which is qualitatively consistent with the Efros-Shklovskii law \cite{efros_coulomb_1975}, suggesting that electron correlations open a gap in the density of states. 
However, the localization length obtained by fitting the temperature dependence exceeds by more than one order of magnitude the one obtained from the magnetic field dependence. We attribute this discrepancy to the presence of a nearby metallic gate, which provides electrostatic screening and thus reduces long-range Coulomb interactions.

\section{Results and discussion}

In this study, a bilayer \mos with dual-gated architecture is employed to study electron transport.
Figure~\ref{fig:Fig1}(a) shows optical images (upper panel) and a schematic side view (lower panel) of the device. 
The fabrication starts by assembling a thin hBN and graphite layers onto a silicon/silicon dioxide chip (\SI{285}{nm} of oxide layer). The graphite serves as the bottom gate, while the hBN is the gate dielectric material. 
The layers are stacked together using a dry-transfer technique (see Refs.~\cite{pisoni_interactions_2018, pisoni_absence_2019, masseroni_electron_2021} for details).
We pattern metallic contacts (Cr/Au: \SI{5}{nm}/\SI{10}{nm}) with standard electron beam lithography and electron beam evaporation. 
The region of the contacts is then cleaned thoroughly with the tip of an atomic force microscope in contact mode. 
This step is crucial to remove the residues of the lithography process.
The bilayer \mos is obtained by mechanically cleaving a bulk \mos crystal from natural sources (SPI supply) and identified by optical contrast, which has proven to be a very reliable method \cite{Ni_graphene_2007, Li_optical_2012, yang_rapid_2013, bing_optical_2018}. 
A second stack with a layer of hBN and the bilayer \mos is assembled, then aligned and deposited on top of the contacts. 
Both exfoliation and assembling take place in a glove box with argon atmosphere ($\mathrm{H_2O,\ O_2}<\SI{0.1}{ppm}$). 
The metallic top gate is patterned with standard lithography processes and covers the entire \mos flake. 
In the final step, we vacuum anneal the sample at $\SI{250}{\degreeCelsius}$ for $\SI{4}{h}$ to improve the contact interface.
All measurements presented in this work (if not explicitly stated otherwise) are performed at a temperature of $\SI{1.3}{K}$ at low frequency ($\sim\SI{30}{Hz}$) and low excitation voltage ($V^\mathrm{rms}_\mathrm{SD}=\SI{100}{\mu V}$) with standard lock-in techniques.  

First, we characterize the contacts to ensure Ohmic behavior and low contact resistance. The dc output characteristics are shown in Fig.~\ref{fig:Fig1}(b) for different top gate voltages ($\vtg$). 
The two-terminal resistance changes from $<\SI{1}{k\Omega}$ to $\sim \SI{1}{M\Omega}$ as a function of $\vtg$. 
The current shows a linear dependence on the applied source-drain voltage $\vsd$ down to the lowest $\vtg$ (inset), indicating a vanishing Schottky barrier. 
The average contact resistance of source and drain contacts (1 and 2) is $< \SI{300}{\Omega}$ for $\vtg\geq \SI{6}{V}$ and does not depend on the applied bottom gate voltage ($\vbg$). 
To the best of our knowledge, this is one of the lowest values reported in the literature and is comparable to bismuth contacts \cite{shen_ultralow_2021}. 
The vanishing Schottky barrier and the low contact resistance allow us to study electron transport at low densities (down to $\SI{1.4E12}{cm^{-2}}$) at low source-drain voltage ($\SI{100}{\mu V}$). 

The four-terminal conductivity ($\sigma$) as a function of $\vtg$ is shown in Fig.~\ref{fig:Fig1}(c). 
We identify two voltage ranges, where the conductivity features a linear dependence on $\vtg$, which differs by the slope.
The two slopes of $\sigma(\vtg)$ yield the mobilities $\sim\SI{1000}{cm^2V^{-1}s^{-1}}$ (dotted line) and $\sim\SI{2400}{cm^2V^{-1}s^{-1}}$ (dashed line).
This specific shape of the four-terminal conductivity is a general property of high-quality single-gated \mos devices [see Fig.~1(c) in Ref.~\cite{lin_determining_2019} and Fig.~1(d) in the supplemental material of Ref.~\cite{pisoni_interactions_2018} for a direct comparison].
The kink marked with the number 2 is related to the population of the upper spin-orbit (SO) split band, as will be demonstrated below. 

Figure~\ref{fig:Fig1}(d) shows the derivative of the conductivity $\mathrm{d}\sigma/\mathrm{d}\vbg$ as a function of $\vtg$ and $\vbg$.
For $\vbg<\SI{0.7}{V}$ the $\vbg$ dependence of the conductivity is monotonic and resembles its $\vtg$ dependence. There are two distinct rates $\mathrm{d}\sigma/\mathrm{d}\vbg$ that are separated by a black dashed line (number 2). This line follows constant density conditions, separating the single band from the two-band regime. From this result, we conclude that the separation between the bands is not displacement field dependent (i.e., the SO gap is not tunable with the electric field). 
In contrast, for $\vbg>\SI{0.7}{V}$  the dependence of the conductivity on $\vbg$ is nonmonotonic. 
The conductivity features a local maximum (number 3) that is almost independent of $\vtg$.
The origin of the negative $\mathrm{d}\sigma/\mathrm{d}\vbg$ is related to the population of the bottom layer and will be discussed in Sec.~\ref{Sec:Band_and_layer_occupation}.

\begin{figure*}
    \centering
    \includegraphics{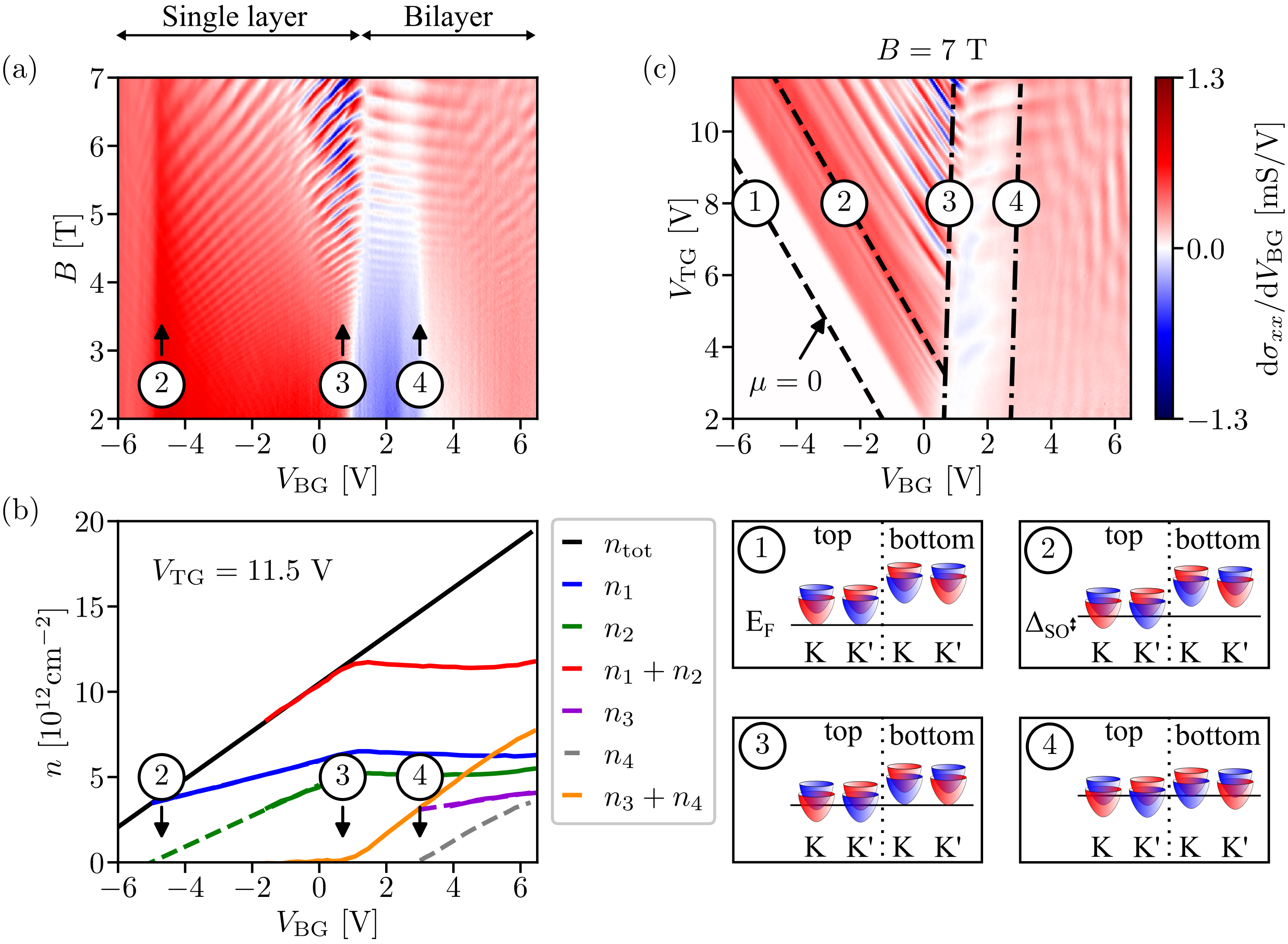}
    \caption{(a) Numerical derivative of the conductivity
    ($\mathrm{d}\sigma_{xx}/\mathrm{d}\vbg$) as a function of $B$ and $\vbg$ for $\vtg=\SI{11.5}{V}$ [for color scale see (c)]. (b) Density as a function of $\vbg$. The total density $n_\mathrm{tot}$ is obtained by Eq.~\eqref{eq:Total_density}, $n_{1,2,3}$ are determined from the Fourier spectrum of the transconductivity in (a) and $n_{4}=n_\mathrm{tot}-n_1-n_2-n_3$. (c) $\mathrm{d}\sigma_{xx}/\mathrm{d}\vbg$ at $B=\SI{7}{T}$ as a function of $\vtg$ and $\vbg$. Lower right panels (1--4) show the schematic of the conduction band and the layer occupancy of the single layer (1 and 2) and bilayer (3 and 4) regimes. The color of the bands encodes the spin polarization.} 
    \label{fig:Fig2}
\end{figure*}

So far we have characterized the field effect transistor at zero magnetic field. 
Understanding the contribution of the bands in multilayer \mos devices is an essential step for interpreting their electronic properties. 
In the following, we will demonstrate that all features observed in the conductivity can be attributed to the population of different bands in the \mos bilayer. 
To verify this hypothesis, we now turn our attention to magnetotransport measurements, from which we determine the electron density and the band occupation.

\subsection{Band and layer occupation}\label{Sec:Band_and_layer_occupation}

The aim of this part is to determine how the different bands are filled with electrons as we change the gate voltages. 
For this purpose we measure the magnetoconductivity $\sigma_{xx}$, applying a perpendicular magnetic field $B$. 
The conductivity $\sigma_{xx}$ is obtained by tensor inversion of the two-dimensional resistivity with components $\rho_{xx}=W V_{3,4}/(I L_2)$ and $\rho_{xy}=V_{3,5}/I$.
From the Shubnikov--de Haas oscillations (SdHO) we determine the density $n_{i}$ of band $i$ as we did in previous works \cite{pisoni_interactions_2018, pisoni_absence_2019, masseroni_electron_2021}.
The total density is given by
\begin{equation}\label{eq:Total_density}
    n_\mathrm{tot} = \frac{1}{e}(C_\mathrm{T} V_\mathrm{TG} + C_\mathrm{B} V_\mathrm{BG}),
\end{equation}
where $C_\mathrm{T}=\SI{146}{nF/cm^{2}}$ and $C_\mathrm{B}=\SI{225}{nF/cm^{2}}$.
The total density is related to the band densities via $n_\mathrm{tot}=\sum_{i} n_{i}$.
We do not determine the total density from the Hall effect, because the contacts extend across most of the conducting channel, leading to an overestimation of the density. 

Figure~\ref{fig:Fig2}(a) shows the derivative of the magnetoconductivity $\mathrm{d}\sigma_{xx}/\mathrm{d}V_\mathrm{BG}$ as a function of $B$ and $\vbg$ for $\vtg=\SI{11.5}{V}$. 
The conductivity features oscillations periodic in $B^{-1}$, as we expect for SdHO.
From the Fourier spectrum of the $\mathrm{d}\sigma_{xx}/\mathrm{d}\vbg(1/B)$ (see Ref.~\cite{pisoni_absence_2019, masseroni_electron_2021} for details) we determine the densities $n_{i}$ shown in Fig.~\ref{fig:Fig2}(b).
As in Ref.~\cite{pisoni_absence_2019}, we attribute a twofold valley degeneracy to each band $i$, which accounts for the two $K$ valleys (see schematics in the lower right panel of Fig.~\ref{fig:Fig2}).
In the regime $\vbg<\SI{0.7}{V}$, where we find only two frequencies in the Fourier spectrum, there is remarkable agreement between the calculated $n_\mathrm{tot}$ and the experimentally defined density $n_{1}+n_{2}$.
In our interpretation, the densities $n_1$ and $n_2$ belong to the top \mos layer.
In the regime $\vbg>\SI{0.7}{V}$, the density $n_{1}+n_{2}$ saturates, because the bottom \mos layer becomes conducting, screening the field effect of the bottom gate.
This behavior is in complete agreement with the behavior found and explained in previous studies \cite{pisoni_absence_2019, masseroni_electron_2021}.

\begin{figure*}
    \centering
    \includegraphics{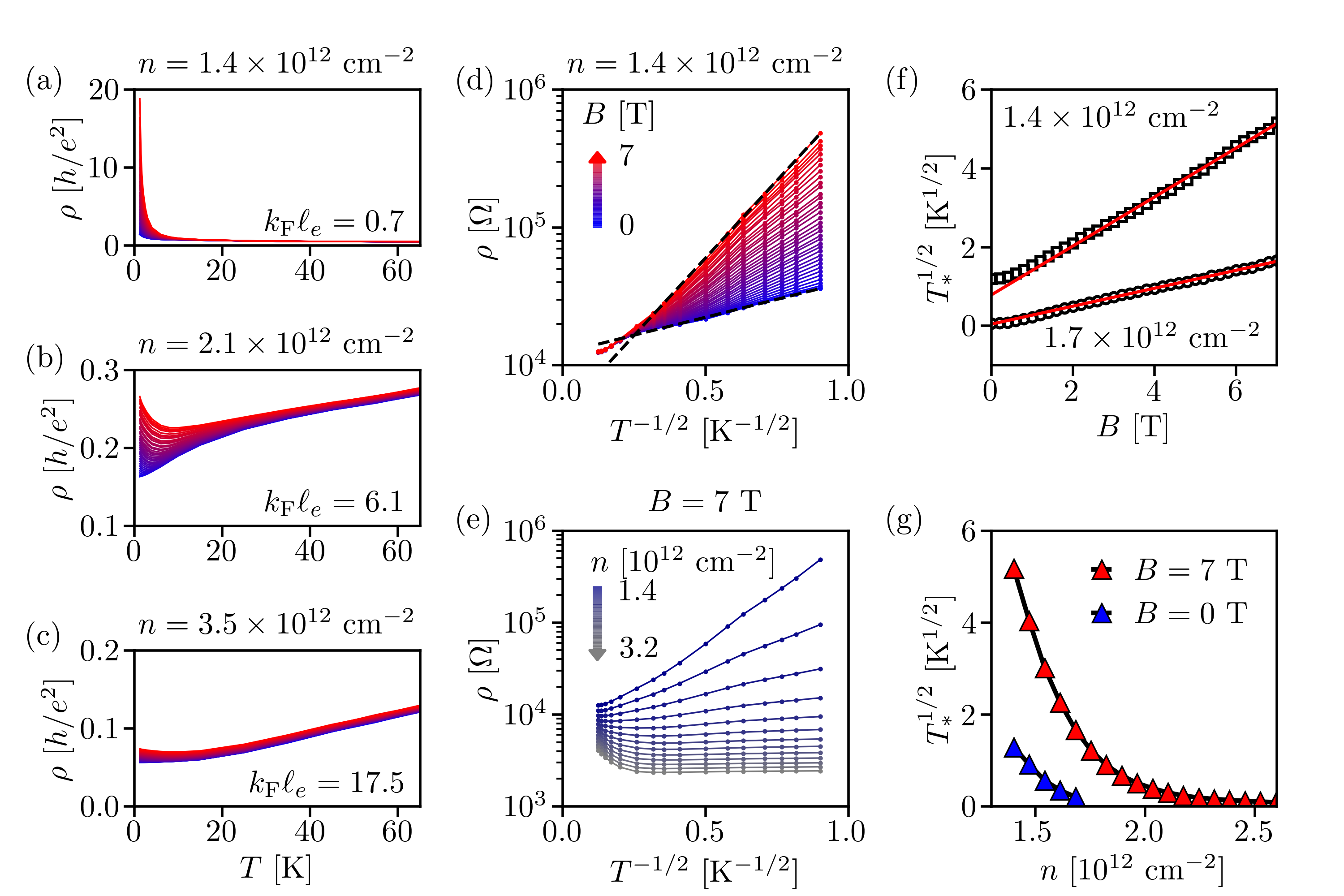}
    \caption{(a--c) Resistivity $\rho$ as a function of temperature $T$ for three electron densities, from $\SIrange{1.4E12}{3.5E12}{cm^{-2}}$, and various magnetic fields. Colorbar in (d). (d) Resistivity  plotted on a logarithmic scale versus $T^{-1/2}$ for $n=\SI{1.4E12}{cm^{-2}}$ and various magnetic fields. The plot shows that the temperature dependence is in agreement with ES law for any magnetic field between 0 and 7~T. (e) Similar to (d) but at $B=\SI{7}{T}$ and various densities. The plot shows the transition from insulating to metallic temperature dependence. (f) The parameter $T^{1/2}_{*}$ is obtained from (d) for two exemplar densities. The solid red lines are linear fits. (g) The parameter $T^{1/2}_{*}$ obtained from (e) for $B=0$ and $B=\SI{7}{T}$.}
    \label{fig:Fig3}
\end{figure*}

The layer occupation can be controlled in dual-gated devices. 
For $\vtg>0$ and $\vbg<0$ only the top layer is occupied by electrons and we can tune between single- and double-band transport. 
The onset of $n_2$ (marked by the number 2) corresponds to the population of the upper SO split band of the top layer (see schematics of the band structure in Fig~\ref{fig:Fig2}).
To support this interpretation we determine the density $n_\mathrm{SO}\approx\SI{3.5E12}{cm^{-2}}$ that is required to start filling the upper SO split band.
In an effective mass approximation, this density corresponds to an energy $\Delta_\mathrm{SO}\sim \SI{14}{meV}$, in agreement with the value reported in Ref.~\cite{pisoni_absence_2019}. 

Figure~\ref{fig:Fig2}(c) shows $\mathrm{d}\sigma_{xx}/\mathrm{d}\vbg$ at $B=\SI{7}{T}$ as a function of the voltage applied to the top and bottom gates.
This measurement defines a phase diagram for the bilayer \mos that is divided into five different regimes by the dashed (dotted) lines.
Each line indicates the onset of a specific band (from 1 to 4 increasing $\vbg$).
The bilayer is tuned from an insulating into a conducting phase, where up to four bands contribute to electron transport.
In the remaining part of this paper, we focus on the narrow single-band transport regime enclosed by the dashed lines in Fig.~\ref{fig:Fig2}(c). 

\subsection{Strong electron localization}

We now turn our attention to the magnetoresistance in the single-band transport regime. 
According to the Drude model, the longitudinal resistance does not depend on the magnetic field. 
However, we observe a non-saturating positive magnetoresistance in the low-density range ($<\SI{3.5E12}{cm^{-2}}$). 
The magnetoresistance increases by more than one order of magnitude at $B=\SI{7}{T}$ and $n_\mathrm{tot}=\SI{1.4E12}{cm^{-2}}$, as shown in Figs.~\ref{fig:Fig3}(a)--\ref{fig:Fig3}(c).
A qualitatively similar positive magnetoresistance was observed in monolayer \mos \cite{schmidt_quantum_2016}, where a negative magnetoresistance (due to weak localization) gradually turned into a positive magnetoresistance as the temperature and density were lowered.
This effect was attributed to the transition from weak localization to weak anti-localization, despite the authors realizing that the shape of the magnetoresistance was not well described by the theory of weak antilocalization. 
Also, in another bilayer \cite{papadopoulos_weak_2019, pisoni_absence_2019} and three-layer \mos devices \cite{masseroni_electron_2021} a similar behavior has been observed but was not further investigated. 
The observation of a positive magnetoresistance in samples prepared and studied in different research groups suggests that there is a common origin for this effect. 
Owing to the relatively low electron mobility in \mos and the large intrinsic defect density \cite{rhodes_disorder_2019}, we consider the role of disorder \cite{ahn_disorder-induced_2022} and electron localization to describe the nonsaturating positive magnetoresistance.

Figures~\ref{fig:Fig3}(a)--\ref{fig:Fig3}(c) show the four-terminal longitudinal resistivity $\rho$ as a function $T$ for three different densities (from $\SIrange{1.4E12}{3.5E12}{cm^{-2}}$) and various magnetic fields. 
At zero magnetic field (violet curve) we observe a metal-insulator transition as we lower the density. 
It is remarkable that this transition occurs at relatively high densities $\sim \SI{1.7E12}{cm^{-2}}$, where the ratio between Coulomb and kinetic energy is still moderate $r_\mathrm{s}=E_\mathrm{C}/E_\mathrm{kin}\sim 7$. 
Therefore, the transition is more likely to be a manifestation of strong electron localization when the Fermi energy approaches the bottom of the conduction band, rather than an insulating state due to correlations. 
At the lowest temperature, we estimate the product ($k_\mathrm{F}\times\ell_e$) between Fermi wave vector ($k_\mathrm{F}$) and electron mean free path ($\ell_e$). 
Strong electron localization occurs for $k_\mathrm{F}\times\ell_e \leq 1$ [as shown in Fig.~\ref{fig:Fig3}(a)], while band conduction occurs for $k_\mathrm{F}\times\ell_e \gg 1$.
Applying a perpendicular magnetic field breaks this condition, inducing a metal-insulator transition even for $k_\mathrm{F}\times\ell_e > 1$ [see Fig.~\ref{fig:Fig3}(b)]. 

The resistivity is plotted on a logarithmic scale as a function of $T^{-1/2}$ in Fig.~\ref{fig:Fig3}(d) for the lowest density and various magnetic fields. 
At low temperatures ($<\SI{10}{K}$) all the curves follow a linear dependence, as predicted by the ES theory.
The slope of the curves increases with increasing magnetic field. Therefore, we define a parameter $T_*(B)$ that depends on the magnetic field, such that the resistivity is described by  
\[\rho (T, B) = \rho_0 \exp\left[ \left(\frac{T_*(B)}{T}\right)^{1/2} \right].  \]
This function seems to capture the temperature dependence of the resistivity for $T<\SI{10}{K}$, while above this temperature the resistance shows a weaker temperature and magnetic field dependence. 
The parameter $T_*^{1/2}$ obtained from the fit is shown in Fig.~\ref{fig:Fig3}(f).
This parameter depends roughly linearly (at least above $B\approx \SI{1}{T}$) on the applied magnetic field and the slope decreases with increasing electron density.
Similarly, we plot the resistivity for different densities [see Fig.~\ref{fig:Fig3}(e)] and extrapolate the parameter $T_{*}^{1/2}$ as a function of densities. 
The density dependence of $T_{*}^{1/2}$ is shown in Fig.~\ref{fig:Fig3}(g), where we see $T_{*}^{1/2}$ increasing rapidly for decreasing density. 

\begin{figure*}
    \centering
    \includegraphics{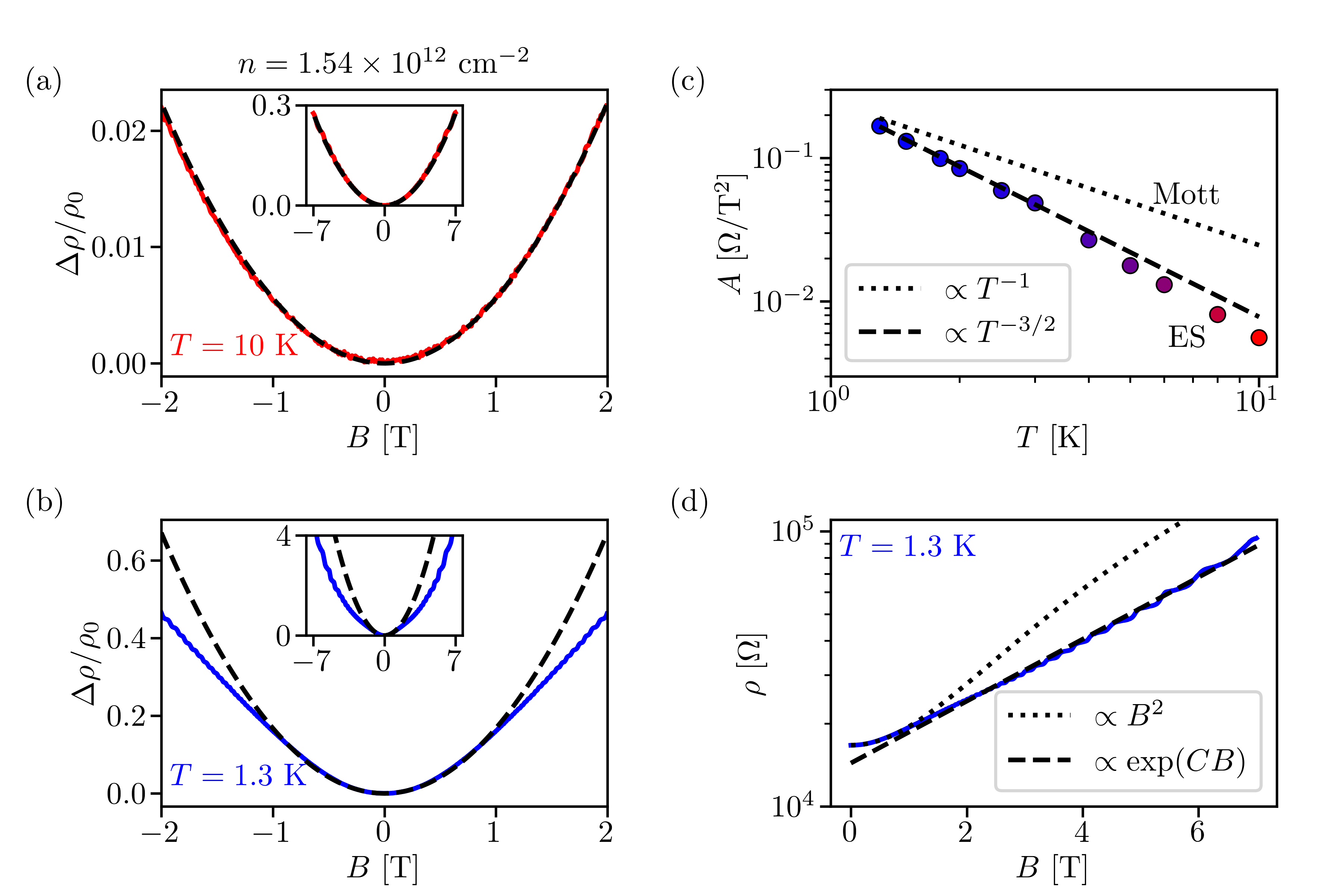}
    \caption{(a and b) Magnetoresistivity $\Delta\rho(B)/\rho(B=0)$ as a function of $B$ for $T=\SI{10}{K}$ and $T=\SI{1.3}{K}$. Inset: same measurement but shown in a larger magnetic field range. (c) The curvature $A(T)$ of $\Delta\rho(B)/\rho(B=0)$ as a function of $T$ on a log-log scale. The dashed (dotted) line shows the expected temperature dependence according to ES (Mott) law. (d) $\rho(B)$ in a logarithmic scale at the lowest temperature ($\SI{1.3}{K}$). The dashed line is a linear fit and serves as a guide for the eyes. The dotted line is the parabolic fit shown in (b). }
    \label{fig:Fig4}
\end{figure*}

We considered other models to describe our data, such as Mott's law ($p=1/3$) and thermally activated nearest-neighbor hopping ($p=1$). 
While thermally activated hopping clearly fails in describing the temperature dependence of the resistivity, we obtain reasonable fits also with Mott's law, which yields a slightly larger least-mean-square error.
Based on the temperature dependence of the resistivity, we cannot reliably distinguish between the two models because the resistivity does not change by several orders of magnitude. 
For this reason, we also consider the role played by the magnetic field.

At finite magnetic fields, the tails of the electron wave functions decay faster and the overlap between the localized wave functions decreases.
This leads to a reduction of the tunneling probability and thus to an increase in the resistance. 
This effect results in an exponential increase of the resistance of the form \cite{shklovskii_electronic_1984}
\[\rho(B) \propto \exp\left(B^{m}\right), \]
where $m$ depends on the range of magnetic field and the assumptions made in the theoretical model \cite{shklovskii_electronic_1984, mikoshiba_strong-field_1962, ioffe_giant_2013}.

First, we focus on the low magnetic field range ($\ell_B \gg a$, $\ell_B$ being the magnetic length and $a$ the localization length), where the effect of $B$ can be treated as a small correction to $T_0$.
The low magnetic field correction of the percolation parameter $\xi$ is given by \cite{nguyen_van_lien_crossovers_1995}
\[ \Delta \xi (B) = \xi(B)-\xi(0) = C_2 \frac{a^4}{\ell_\mathrm{B}^4} \left(\frac{T_0}{T}\right)^{3/2} = A(T) B^2,\]
where $C_2=0.002$ is a numerical parameter, and $\ell_B=\sqrt{h/eB}$.
This expression yields the magnetoresistivity
\begin{equation}\label{eq:Resistivity_VRH_B}
    \frac{\Delta\rho(B)}{\rho(0)} = \exp\left[ A(T) B^2\right]-1\approx A(T) B^2,
\end{equation}
which can be expanded in a quadratic expression for $A(T) B^2 << 1$.

Figures~\ref{fig:Fig4}(a) and \ref{fig:Fig4}(b) show the low magnetic field range of the resistivity at the density $\SI{1.54E12}{cm^{-2}}$ (i.e., below $n_\mathrm{c}$) and for two exemplary temperatures. 
At $T=\SI{10}{K}$ the resistivity shows a quadratic dependence on $B$, as predicted by Eq.~\eqref{eq:Resistivity_VRH_B}.
At this temperature, the model fits the data well in the entire magnetic field range probed in our experiments, as shown by the inset of Fig.~\ref{fig:Fig4}(a).
The range of magnetic field for which this model is able to describe the data shrinks with lowering the temperature because the parameter $\Delta \xi(B)$ grows with lowering the temperature ($\propto T^{-3/2}$), thus limiting the magnetic field range for which our approximation is valid.
Therefore, it is not surprising that at $T=\SI{1.3}{K}$ [Fig.~\ref{fig:Fig4}(b)] the model deviates from the data at $B_\mathrm{c}\approx\SI{1}{T}$ where $\Delta\xi/\xi_0\approx 0.5$ becomes a large correction.

Fitting the magnetoresistance at different temperatures provides the temperature dependence of the curvature, which is the only fitting parameter. 
The result of the fit is shown in Fig.~\ref{fig:Fig4}(c) on a log-log scale for the density $\SI{1.54E12}{cm^{-2}}$.
The data follow the temperature dependence $T^{-(1.6 \pm 0.1)}$.
For comparison, we show the temperature dependence ($T^{-1}$) predicted by the Mott theory, which clearly deviates from the trend of our data, while the data are in good agreement with ES theory ($T^{-3/2}$). 
Based on this observation, we conclude that, in ``diluted" \mos, long-range Coulomb interactions lead to the formation of a gap in the density of states at the Fermi energy. This conclusion is further supported by the interaction parameter $r_\mathrm{s}\sim 7$, which is modest but indicates that the Coulomb energy is significantly larger than the kinetic energy of free electrons.

At this point, we would like to discuss our quantitative results. 
By fitting the data with ES law we can estimate the localization length $a$ according to 
\begin{equation}\label{eq:T_0}
    k_\mathrm{B} T_0 = C_1 \frac{e^2}{4\pi\epsilon_0 \epsilon_r a},
\end{equation}
where $\epsilon_r\approx 7$ is the relative  dielectric constant of \mos, and $C_1=6.2$ is a numerical constant \cite{nguyen_van_lien_crossovers_1995}. 
This equation, however, provides localization lengths of a few micro\-meters, which is surprisingly large considering the inter-particle spacing ($n^{-1/2}\sim \SI{10}{nm}$). 
To verify the validity of this result, we estimate the localization length from the curvature of the magnetoresistance and compare the two results. 
By inserting $T_0$ (obtained from the temperature dependence) in the definition of $A(T)$ we obtain $a\approx \SI{100}{nm}$ at a density of $\SI{1.54E12}{cm^{-2}}$.
This result differs by more than one order of magnitude from the one obtained by Eq.~\eqref{eq:T_0}.

Here, we offer an argument that might explain the discrepancy between these two results. 
We note that in our device the metallic gate is only separated from the 2D electron system by a thin hBN layer ($d=\SI{13}{nm}$).
This distance is comparable to the interparticle distance $n^{-1/2}\sim \SI{10}{nm}$.
As a consequence, the metallic gate screens Coulomb interactions at large distances ($r\gg d$), where the Coulomb potential becomes dipole-like ($\propto r^{-3}$).
The presence of the metal plate partially suppresses the Coulomb gap due to its screening effect \cite{aleiner_effect_1994, cuevas_electrode_1994, pikus_coulomb_1995, hu_screening_1995, van_keuls_screening_1997, ho_reduced_2010}.
For small hopping distances the density of available states still has a gap, while for long hopping distances the density of states is constant. 
At lower temperatures, long-distance hopping becomes favorable and VRH is determined by the constant density of states (like in the Mott VRH).
Therefore, it is expected that in the presence of a nearby metallic gate, there is a critical temperature
$$T_\mathrm{c}=\frac{e^2}{4\pi\varepsilon_0\varepsilon_r k_\mathrm{B}} \frac{a}{d^2}$$
below which the system shows a transition from ES to the ``screened" Mott VRH \cite{hu_screening_1995}.
This VRH is different from the usual Mott VRH, which is characterized by a different density of states.

We estimate this temperature to be around $\SI{20}{K}$ by assuming $a\approx \SI{100}{nm}$, placing our system just below the transition temperature. 
In contrast, the hopping distance
$$r_\mathrm{h} \sim \frac{1}{4} a \left(\frac{T_0}{T}\right)^{1/2} \sim \SI{10}{nm}$$
is of the order of $d$, thus still in the Coulomb gap regime.
Therefore, it is not clear from these characteristic quantities in which regime our sample is. 
As we argued above, we cannot, based on the temperature dependence, clearly distinguish between ES and Mott VRH.
Under this condition, Eq.~\eqref{eq:T_0} may not be valid, as we are close to the transition between the two hopping transport regimes. 
On the other hand, the magnetic field promotes electron localization and reduces the probability of long-range hopping (less overlap of wave function tails). 
As shown in Fig.~\ref{fig:Fig3}(f), the magnetic field dependence of the fitting parameter $T_{*}(B)\propto B^2$ is compatible with the wave-function shrinking effect. 
In addition, the curvature $A(T)$ clearly establishes that at finite field VRH follows ES theory. 
For this reason, we consider our estimation of the localization length from the analysis shown in Fig.~\ref{fig:Fig4} to be valid, from which we estimate $a\sim\SI{100}{nm}$.

Finally, we consider the asymptotic limit at high magnetic fields ($a \gg \ell_B$). 
The typical hopping distance is strongly reduced by the effect of the magnetic field and the typical hopping energy increases.
As a consequence, the Coulomb gap might be negligible at high $B$-fields, as proposed by Nguyen \cite{nguyen_van_lien_crossovers_1995}. 
The resistivity is expected to follow the asymptotic behavior described by Eq.~\eqref{eq:Asymptotic_rho},
\begin{equation}\label{eq:Asymptotic_rho}
    \frac{\Delta \rho(B)}{\rho(0)} \propto  \exp(C\sqrt{B}),
\end{equation}
with $C\propto T^{-1/2}$ being a temperature-dependent coefficient. 
This equation captures the temperature dependence of the resistivity at high magnetic fields. 
However, the resistivity seems to follow $\exp(CB)$ instead of Eq.~\eqref{eq:Asymptotic_rho}, as shown in Fig.~\ref{fig:Fig4}(d).
Given the divergent susceptibility of \mos at low density \cite{lin_determining_2019}, we speculate that the spin polarization might contribute in increasing the magnetoresistance.
As it was reported in Ref.~\cite{meir_universal_1996} the mechanism that provides an enhanced magnetoresistance is related to the blocking of hopping due to spin polarization.
Disentangling spin from orbital effects requires in-plane magnetic fields. However, the in-plane magnetic field does not couple to the electron spins in the $K$ valleys of \mos, because the spin is locked out-of-plane due to SO coupling \cite{movva_density-dependent_2017, lin_determining_2019}.

\section{Conclusion}

The low-temperature resistivity of bilayer \mos undergoes a transition from metallic to insulating temperature dependence at a critical density $n_\mathrm{c}\approx\SI{1.7E12}{cm^{-2}}$. This density is one order of magnitude larger than in silicon metal-oxide-semiconductor field-effect transistors \cite{tracy_observation_2009} and three orders of magnitude larger than in GaAs quantum wells \cite{simmons_metal-insulator_1998, manfra_transport_2007}. 
We attribute this transition to a disorder-induced transition in agreement with other metal-insulator-transition observed in disordered materials \cite{tracy_observation_2009}. 
In fact, our observation is in line with the proposal of Klapwijk and Das Sarma \cite{klapwijk_few_1999}, which states that the transition should be observed when the electron density is close to a few electrons per ionized impurity. 
The ionized impurities may be sulfur vacancies, which are known to have an inhomogeneous distribution with an average density of $\SI{1E12}{}-\SI{1E13}{cm^{-2}}$ \cite{vancso_intrinsic_2016}, thus the same order of magnitude as $n_\mathrm{c}$ in our sample.

In the insulating phase, the resistance drops exponentially with increasing temperature, compatible with variable-range hopping.  The limited temperature range considered in our experiments does not provide a conclusive distinction between Mott and Efros-Shklovskii laws. 
On the other end, the magnetic field dependence at low fields closely follows an Efros-Shklovskii law.
The Coulomb gap is likely to appear in the density of states of \mos, as the interaction parameter ($r_\mathrm{s}\sim 7$) suggests that Coulomb energy is the dominant energy scale.
However, the presence of a nearby metallic gate could contribute to the suppression of the gap in the density of state at large distances ($r_h>2d$) \cite{cuevas_electrode_1994, hu_screening_1995, van_keuls_screening_1997, ho_reduced_2010}.
Despite the screening effect, we observe the presence of the Coulomb gap by applying a perpendicular magnetic field. 
We interpret this result based on the wave function shrinking, which reduces long distant hopping.

\section*{Acknowledgments}
We thank Boris Shklovskii, Hadrien Duprez, and David Kealhofer for fruitful discussions.
We thank Peter Märki, Thomas Bähler, as well as the FIRST staff for their technical support.
We acknowledge support from the European Graphene Flagship Core3 Project, Swiss National Science Foundation via NCCR Quantum Science, and H2020 European Research Council (ERC) Synergy Grant under Grant Agreement 95154.
K.W. and T.T. acknowledge support from JSPS KAKENHI (Grants No. 19H05790, No. 20H00354, and No. 21H05233).

%

\end{document}